\documentclass[aps
floats,prd,nofootinbib,superscriptaddress,10pt]{revtex4-1}
\usepackage[utf8]{inputenc}
\usepackage{bm}
\usepackage{amsmath}
\usepackage{comment}
\usepackage{caption}
\usepackage{color}

\usepackage{graphicx,amsmath,amsfonts,amssymb
}

\begin{document}
\title{Reheating through the Higgs amplified by spinodal instabilities\\and 
gravitational creation of gravitons}

\author{Tomohiro Nakama}

\affiliation{Department of Physics and Astronomy, Johns Hopkins
     University, 3400 N.\ Charles St., Baltimore, MD 21218, USA}

\affiliation{Institute for Advanced Study, The Hong Kong University of Science and Technology, Clear Water Bay, Kowloon, Hong Kong, P.R.China}

\author{Jun'ichi Yokoyama}

\affiliation{Research Center for the Early Universe (RESCEU),
Graduate School of Science, The University of Tokyo, Tokyo 113-0033, Japan}

\affiliation{Department of Physics, Graduate School of Science, The University of Tokyo,
Tokyo, 113-0033, Japan}

\affiliation{Kavli Institute for the Physics and Mathematics of the Universe (Kavli IPMU),
UTIAS, WPI, The University of Tokyo, Kashiwa, Chiba, 277-8568, Japan}

\begin{abstract}
It is shown that a positive non-minimal coupling of the Higgs field
to gravity can solve the two problems in inflation models in which
 postinflationary universe is dominated by an energy with stiff equation 
of state such as a kination, namely, overproduction of gravitons in
gravitational reheating scenario, and overproduction of curvature
perturbation from Higgs condensation.  Furthermore, we argue that 
the non-minimal coupling parameter can be constrained more stringently
with the progress in observations of large-scale structure and cosmic
microwave background. 
\end{abstract}

\maketitle

\section{Introduction}
If the equation of state parameter $w$ is larger than $1/3$ after inflation \cite{Sato:2015dga},
 the energy density of the inflaton decreases more quickly than radiation. For instance, the universe can be dominated by the kinetic energy of the inflaton ($kination$) after 
k-inflation \cite{ArmendarizPicon:1999rj}, G-inflation \cite{Kobayashi:2010cm,Kobayashi:2011nu}, or quintessential 
inflation \cite{Peebles:1998qn}, when $w=1$ and the energy density
 decays 
as $a^{-6}$ where $a$ is the cosmic scale factor. 
A relatively large $w$ may also be realized by an inflaton oscillating
 around 
a minimum of its potential if it is steeper than quartic \cite{Ford:1986sy}. 

 In these models reheating is supposed to take place due to gravitational 
particle production.  Since spinor and gauge fields are
conformally 
invariant without mass terms, production
of minimally coupled massless scalar fields has been discussed as a
source 
of radiation in these 
models  \cite{Ford:1986sy} and used in the literatures 
\cite{ArmendarizPicon:1999rj,Kobayashi:2010cm,Peebles:1998qn}.  
Since each polarization mode of gravitons satisfies the same equation
of 
motion to the linear order, 
they are produced twice as much as a massless minimally coupled scalar
field.  
Hence if this is the only mechanism
of reheating after inflation, gravitons would be overproduced relative to radiation unless we introduce sufficiently
many light bosons.  Thus we should seek for other sources of entropy after inflation.

One of the candidates is the standard Higgs field as it may acquire a 
condensation of long-wave quantum fluctuations which are generated if it
is 
minimally coupled to gravity so that its  effective mass is much smaller
than the Hubble parameter during inflation \cite{Starobinsky:1994bd}. 
If it has a typical amplitude around the root-mean-square in the domain 
corresponding to our Universe,
its energy density is of the same order of radiation energy created by 
gravitational particle production 
just after inflation \cite{Kunimitsu:2012xx}.
However, since its amplitude remains constant until the Hubble parameter
decreases 
to its effective mass,  its energy
density would surpass that of both the remnant inflaton and radiation
created 
gravitationally.  Hence the Universe is 
reheated by the Higgs condensation which mainly decays to gauge bosons \cite{Figueroa:2015rqa,Enqvist:2015sua}.
It has been shown, however, such a scenario does not yield sensible cosmology
 because the Higgs condensation has too large fluctuations
which generate too large curvature perturbation \cite{Kunimitsu:2012xx}.
  
In this paper we argue that if the Higgs field is non-minimally coupled
to the 
scalar curvature, we can not only solve the
overproduction problems of both curvature fluctuations and gravitons
but 
also find interesting observational constraints
due to relic gravitons.  Indeed if the Higgs field has a large enough 
positive non-minimal coupling such as the conformal coupling
$\xi=1/6$, then it has a large enough effective mass 
$m^2_{\rm eff}=\xi R =12\xi H^2$ during inflation 
and no long-wavelength perturbations are generated  
\cite{Kamada:2014ufa,Herranen:2015ima,Figueroa:2016dsc}. 
This additional mass term, however, becomes negative after
 inflation if $w>1/3$ then \cite{Figueroa:2016dsc}, and consequently
 it experiences spinodal instabilities
 \cite{Cormier:1999ia,Albrecht:2014sea} 
shortly after inflation, which determines the energy density of the 
Higgs field and hence its decay products. This can provide a more efficient
source of reheating than gravitational particle production. Then 
contribution of gravitons to the radiation can be
sufficiently small depending on the value of the non-minimal coupling.  
Since the presence of graviton radiation  affects the observed cosmic 
microwave background and the structures of the universe 
\cite{Smith:2006nka,Sendra:2012wh}, this scenario can be 
probed by these observations and we can obtain
constraints on the non-minimal coupling. See also 
Ref. \cite{Artymowski:2017pua} for a recent discussion on a different 
aspect of gravitational reheating.

The rest of the paper is organized as follows.
In the next section, we introduce a simple model of a transition from a de-Sitter phase to a kination, which we use as an example, and then discuss gravitational reheating and creation of gravitons at the transition. Then we discuss spinodal instabilities of the Higgs at the transition in Sec. III.
It is convenient to express the energy density of gravitons as an effective, additional contribution to the number of neutrino species \cite{Smith:2006nka,Sendra:2012wh}, which we denote by $N_{\mathrm{eff,GW}}$, to relate it to observations of the cosmic microwave background and the structures of the universe. We present this quantity for several combinations of parameter values in Sec. IV, with comparison to existing constraints and expected, future sensitivities. We conclude in Sec. V.

\section{Gravitational reheating and production of gravitons}
In order to discuss gravitational reheating,  let us take the Hubble parameter to be constant during inflation with the scale factor $a=\exp[H(t-t_0)]$, where $t_0$ is chosen  to be the moment when the spacetime starts to deviate from  de-Sitter expansion toward the end of inflation. 
We denote the conformal time by $\eta$, and
take $\eta=\eta_0$ at $t=t_0$.  Then, 
$(\eta_0-\eta)H=a(t)^{-1}-1.$ Well before $\eta_0$, we have $\eta\simeq-1/(aH). $ 
Let us consider the case the inflaton's energy density, $\rho_{\mathrm{inf}}$, is dominated
by its kinetic energy after inflation so that we find $\rho_{\mathrm{inf}}\propto a^{-6}$, 
and hence
$H(t)\propto a^{-3}, a\propto t^{1/3}$, and $\eta\propto t^{2/3}\propto a^{2}.$

Following Ref. \cite{Kunimitsu:2012xx} we introduce
$f(H\eta)\equiv a^2(\eta)$ and $x=H\eta$, and also normalize $\eta$ so that $H\eta_0=-1$. 
Similarly to Ref. \cite{Kunimitsu:2012xx}, we consider the following transition from a de-Sitter phase to a kination:
\begin{equation}
    f(x) = 
\begin{cases}
    1/x^2  & (x<-1) \\ 
   a_0+a_1x+a_2x^2+a_3x^3+a_4x^4+a_5x^5 & (-1<x<-1+x_0)  \\
   b_0(x+b_1)  & (-1+x_0<x)
\end{cases}\label{eqzero}
\end{equation}
Here, $x_0$ is a parameter describing how rapid the transition from a de-Sitter phase to a kination is,
and the coefficients are determined by requiring $f, f', f'', f^{(3)}$
to be continuous both at $x=-1$
and $ -1+x_0$. 
Then, $\tilde{V}'$, introduced shortly, and the Ricci scalar are continuous throughout the transition regime. 
Here the Ricci scalar is given by
\begin{equation}
R=6\left[\frac{\ddot{a}(t)}{a(t)}+
\frac{\dot{a}^2(t)}{a^2(t)}\right]
=6\frac{a''(\eta)}{a^3(\eta)}, \label{1}
\end{equation}
where the prime denotes derivative with respect to $\eta$.

The energy density of minimally coupled massless scalar particles produced by
the above change of the cosmic expansion law is given by \cite{Kunimitsu:2012xx}
\begin{equation}
\rho_r=\frac{I\cdot H^4}{128\pi^2a^4},\label{rhor}
\end{equation}
where
\begin{equation}
I=-\int_{-\infty}^xdx_1\int_{-\infty}^xdx_2\ln(|x_1-x_2|)\tilde{V}'(x_1)\tilde{V}'(x_2)
=-2\int_{-\infty}^xdx_1\int_{-\infty}^{x_1}dx_2\ln(x_1-x_2)\tilde{V}'(x_1)\tilde{V}'(x_2),
\end{equation}
\begin{equation}
\tilde{V}(x)=\frac{f''f-(f')^2/2}{f^2}.
\end{equation}
The upper limit of the integration above, $x$, should be taken sufficiently larger than the end of the transition at $x=-1+x_0$. 
Numerical integrations revealed $I\simeq 50x_0^{-0.262}$ with 
at least for $0.1<x_0<1$.
This radiation energy surpasses that of kination at $a=a_{\mathrm{end}}\sqrt{32/3}\pi M_p/H\equiv a_R$ 
and the reheating temperature is given by $T_R=M_p[90/\pi^2g_*(T_R)]^{1/4}(32\pi^2/3)^{-3/4}(\Lambda_{\mathrm{COBE}}/M_p)^2(r/0.01)=4\times 10^6(r/0.01)\mathrm{GeV} $ as a function of the 
tensor-to-scalar ratio $r$, where $M_p=2.435\times 10^{18}\mathrm{GeV}$ is the reduced Planck mass, $\Lambda_{{\mathrm{COBE}}}=2.54\times 10^{13}\mathrm{GeV}$ and $g_*(T_R)$ is set to 106.75 (see Ref. \cite{Artymowski:2017pua} for more details), if the Universe is reheated by a minimally-coupled massless scalar field created gravitationally, instead of by the Higgs amplified by spinodal instabilities discussed in the next section.

Since the graviton satisfies the same equation of motion as a massless minimally coupled scalar field, its
 energy density  is twice as large as that given by
  Eq. (\ref{rhor}), reflecting their two polarization states \cite{Ford:1986sy}.

\section{Spinodal instabilities of the Higgs}
We consider the growth of the real and neutral
component of the Higgs field:
\begin{equation}
{\cal L}_\phi=\sqrt{-g}\left(-\frac{1}{2}\partial_\mu\phi\partial^\mu\phi-\frac{1}{2}m^2\phi^2-\frac{1}{4}\lambda\phi^4-\frac{1}{2}\xi R\phi^2\right),
\end{equation}
where we have included the mass term for later convenience but omitted
coupling to other fields including gauge fields. 
The equation of motion for $\phi$ reads
\begin{equation}
\ddot{\phi}+3H\dot{\phi}-a^{-2}\Delta\phi+m^2\phi+\lambda \phi^3+\xi R\phi=0.\label{eom}
\end{equation}
If we take $\xi \gg 1$, then $\phi$ has a large tachyonic mass after
inflation
as $R$ becomes negative.  Then $\phi$ would soon settle down to a
minimum $\phi^2=\xi |R|/\lambda \equiv \phi_m^2$ which is the situation
considered in \cite{Figueroa:2016dsc}.  Here we are interested in
the case $\xi \sim 1$, where  $\phi$ does not
grow  rapidly enough to relax to the minimum $\phi_m$ before $|R|$
declines.  In such a situation we can adequately describe the dynamics
of $\phi$ using the Hartree (Gaussian) approximation to take 
$\phi^3\simeq 3\langle\phi^2\rangle\phi$ \cite{Cormier:1999ia,Albrecht:2014sea}, where 
\begin{equation}
\langle\phi(\bm{x},t)^2\rangle =\int\frac{dk}{k}{\cal P}(k,t),\label{power}
\end{equation}
with
\begin{equation}
\phi(\bm{x},t)=\int\frac{d^3 {k}}{(2\pi)^{3/2}}\phi(\bm{k},t)e^{i\bm{k}\cdot \bm{x}},\quad
\langle\phi(\bm{k},t)\phi^*(\bm{k}',t)\rangle
=\frac{2\pi^2}{k^{3}}\delta(\bm{k}-\bm{k}'){\cal P}(k,t).
\end{equation}
It turns out that the $\lambda$ term is unimportant for parameter values
we consider for which the effect of created gravitons is (potentially)
observable (see TABLE 1 in \S IV).

After inflation, the Ricci scalar becomes negative, and the Higgs starts to grow, but the absolute magnitude of the Ricci scalar decays rapidly during a kination ($|R|\propto a^{-6}$), and as a result this growth is soon shut off.
We evaluate the energy density stored in  the Higgs field  
when the Ricci scalar has sufficiently decayed and hence
spinodal instabilities have terminated. 
Let us write the energy density of $\phi$ as (neglecting the $\xi$ term)
\begin{equation}
\rho_{\mathrm{Higgs}}=\rho_K+\rho_V+\rho_{\mathrm{grad}}, \label{rho}
\end{equation}
where
\begin{equation}
\rho_K=\frac{1}{2}\dot{\phi}^2
=\frac{1}{2}\int\frac{dk}{k}{\cal P}_{\dot{\phi}}(k,t) \label{10}
\end{equation}
with
\begin{equation}
\langle\dot{\phi}(\bm{k},t)\dot{\phi}^*(\bm{k}',t)\rangle
=\frac{2\pi^2}{k^{3}}\delta(\bm{k}-\bm{k}'){\cal P}_{\dot{\phi}}(k,t),\label{11}
\end{equation}
\begin{equation}
\rho_V=\frac{1}{4}\lambda \phi^4
\simeq\frac{1}{4}\lambda\cdot 3\langle\phi^2(\bm{x},t)\rangle^2, 
\label{12}
\end{equation}
and
\begin{equation}
\rho_{\mathrm{grad}}=\frac{1}{2}\frac{(\nabla
 \phi)^2}{a^2}=\frac{1}{2a^2}\int\frac{dk}{k}k^2{\cal P}(k,t). \label{13}
\end{equation}

Introducing a conformally rescaled field variable, $\chi=a\phi$,  the first two terms of Eq. (\ref{eom}) can be rewritten as
\begin{equation}
\frac{1}{a(\eta)}\partial_\eta\left(\frac{1}{a(\eta)}\partial_\eta\left[\frac{\chi(\bm{x},\eta)}{a(\eta)}\right]\right)
+3H\frac{1}{a(\eta)}\partial_\eta\left[\frac{\chi(\bm{x},\eta)}{a(\eta)}\right]=\frac{a(\eta)\chi''(\bm{x},\eta)-a''(\eta)\chi(\bm{x},\eta)}{a(\eta)^4}
=\frac{\chi''}{a^3}-\frac{R\chi}{6a}. \label{14}
\end{equation}
Hence, Eq. (\ref{eom}) becomes
\begin{equation}
\chi''-\Delta \chi+[a^2m^2+3\lambda \langle\chi^2\rangle+a^2(\xi-1/6)R]\chi=0.
\label{15}
\end{equation}
We introduce the mode function $\chi_k(\eta)$ by rewriting each Fourier mode as 
$\chi(\bm{k},\eta)=\chi_k(\eta)a(\bm{k})+\chi_k^*(\eta)a^\dagger(\bm{k})$, where $a$ and $a^\dagger$ satisfy $[a(\bm{k}),a^\dagger(\bm{k}')]=\delta(\bm{k}-\bm{k}')$.
This leads to ${\cal P}_\chi(k,\eta)=k^3|\chi_k(\eta)|^2/2\pi^2$.
The equation of motion for $\chi_k(\eta)$  is 
\begin{equation}
\chi_k''+{\cal M}^2(k,\eta)\chi_k=0,\label{fourier}
\end{equation}
where 
\begin{equation}
{\cal
 M}^2(k,\eta)=k^2+a^2m^2+3\lambda\langle\chi^2\rangle+a^2(\xi-1/6)R.
\label{f}
\end{equation}
Since $m^2$ is presumably much smaller than the scale of inflation and reheating
we neglect it hereafter.  Furthermore, as
 the Ricci scalar is constant during  de-Sitter phase, 
the initial conditions for $\chi$ and $\chi'$, when $\langle\chi^2\rangle$ is small, can be provided by \cite{Bunch:1978yq,Vilenkin:1982wt}
\begin{equation}
\chi_k(\eta)=\exp\left[i\left(\nu+\frac{1}{2}\right)\frac{\pi}{2}\right]
\sqrt{\frac{\pi}{4k}}\sqrt{-k\eta}\,H_\nu^{(1)}(-k\eta), \label{18}
\end{equation}
where 
\begin{equation}
\nu=\sqrt{\frac{9}{4}-\frac{m_{\mathrm{eff}}^2}{H^2}},\quad m_{\mathrm{eff}}^2=12\xi H^2,
\end{equation}
and $H_\nu^{(1)}$ is the Hankel function. 

During a de-Sitter phase, $R=12H^2$, and shortly after $\eta_0$, $R$ decreases rapidly to become negative. After taking a negative minimum value $R_m$ at $\eta=\eta_m$, $|R|$ decays rapidly ($R\propto a^{-6}$) during  kination. The modes satisfying $k^2\lesssim k_M^2\equiv -a^2(\eta_m)(\xi-1/6)R_m$ grow during a short interval after $\eta_0$ due to the last term of Eq. (\ref{f}).  
We use $k_M$ as the upper limit of the $k-$integrations. First we solve
for time evolution of Higgs fluctuations in Fourier space solving
Eq.\ (\ref{15}), with the initial conditions provided using Eq.\ (\ref{18}), at the
moment sufficiently before the end of the de-Sitter phase. To solve
Eq.\ (\ref{fourier}), the time evolution of the Ricci scalar is needed, which can be
obtained using Eq.\ (\ref{1}), assuming the model for a smooth transition from
a de-Sitter phase to a kination phase outlined in  Eq.\ (\ref{eqzero}).  The term $\langle \chi^2\rangle$ in Eq. (\ref{f}) can be computed using Eq. (\ref{power}) at each time step.
The energy density of the Higgs field well after spinodal instability 
terminates can then be determined using Eqs.\ (\ref{rho}), (\ref{10}), 
(\ref{12}) and (\ref{13}). 
On the other hand, the energy density of gravitons generated at the 
transition can be calculated using Eq.\ (\ref{rhor}) for the same transition
model, 
noting that the numerical factor $I$ there is determined by $x_0$, 
which specifies how sudden the transition is. The ratio between these 
two energy densities determines observability of gravitational waves, 
as discussed in the next section. 

\begin{figure}[htbp]
  \begin{center}
    \includegraphics[clip,width=7.0cm]{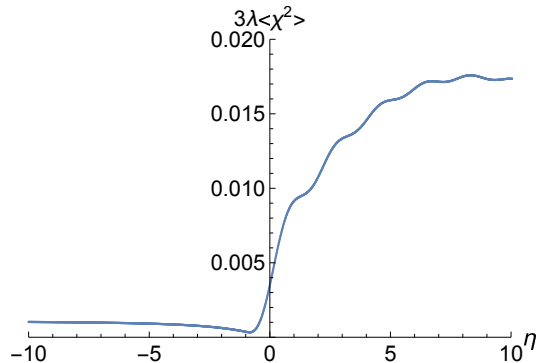}
    \caption{Time evolution of the third term of Eq. (\ref{f}), for $(x_0,\xi,\lambda)=(0.5,1,0.01)$ in the Hubble unit $H=1$.}
    \label{self}
  \end{center}
\end{figure}

\begin{figure}[htbp]
  \begin{center}
    \includegraphics[clip,width=7.0cm]{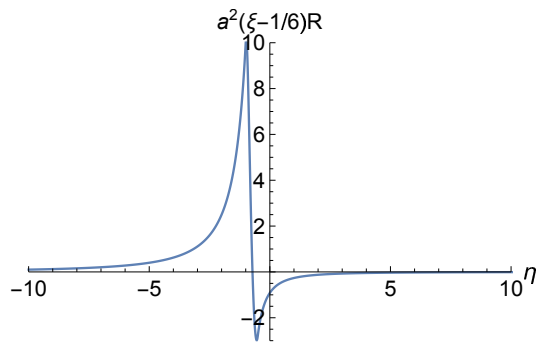}
    \caption{Time evolution of the fourth term of Eq. (\ref{f}), for $(x_0,\xi,\lambda)=(0.5,1,0.01)$ in the Hubble unit $H=1$.}
    \label{source}
  \end{center}
\end{figure}

Time evolution of the third and fourth terms of Eq. (\ref{f}) is shown in Figs. \ref{self} and \ref{source}, for $(x_0,\xi,\lambda)=(0.5,1,0.01)$. In this case, we find $k_M\simeq 1.7 H$, and these show the self-interaction term is negligible even after the instability growth. The fourth term turns out to be $\simeq -0.018 H^2$, whereas the third term is $\simeq 0.017 H^2$ at $\eta=7H^{-1}$ when the latter saturates. 
\section{The relation between the model parameters and $N_{\mathrm{eff,GW}}$}
We calculate $\rho_{\mathrm{Higgs}}/\rho_{\mathrm{GW}}$ at a moment $t_{\mathrm{late}}$ sufficiently later than the end of the transition at $x=-1+x_0$ so that the Higgs field is oscillating with an effectively  quartic potential
but before its decay, when its energy density
 simply decreases as $\propto a^{-4}$ in the same way
as radiation, gradually produced by the Higgs decay.  After the decay of the Higgs, which we assume as a dominant mechanism of reheating, a thermal bath of Standard Model particles is established with energy density $\rho_{\mathrm{rad}}$ at $t=t_{\mathrm{th}}$. Then, we find
\begin{equation}
\frac{\rho_{\mathrm{Higgs}}}{\rho_{\mathrm{GW}}}\bigg|_{t_{\mathrm{late}}}=
\frac{\rho_{\mathrm{rad}}}{\rho_{\mathrm{GW}}}\bigg|_{t_{\mathrm{th}}}. \label{relation}
\end{equation}
After $t_{\mathrm{th}}$, the total energy density $\rho_{\mathrm{tot}}$ is written as
\begin{equation}
\rho_{\mathrm{tot}}
=\rho_{\mathrm{rad}}+\rho_{\mathrm{GW}}
=\frac{\pi^2}{30}g_*T^4+\rho_{\mathrm{GW}},
\end{equation}
where $g_*$ is the number of effective degrees of freedom. We assume the thermalization occurs sufficiently early, when $g_*=106.75$ \cite{Husdal:2016haj}. While $\rho_{\mathrm{GW}}\propto a^{-4}$, the energy density of the radiation of standard particles $\rho_{\mathrm{rad}}$ behaves according to the conservation of entropy: $\rho_{\mathrm{rad}}\propto g_*^{-1/3}a^{-4}$ \cite{Nakama:2016enz}, hence, $\rho_{\mathrm{GW}}/\rho_{\mathrm{rad}}\propto g_*^{1/3}$.\footnote{Here, the difference in the effective degrees of freedom defined in terms of the energy density $g_*$ and the entropy density $g_s$ is neglected, since it is always insignificant ($g_*=3.363$ versus $g_s=3.909$ after the annihilation of $e^+e^-$ \cite{Husdal:2016haj}). For simplicity, we set $g_*=g_s=3.5$, which would be a good approximation given the weak dependence on the degrees of freedom. At the time of the big bang nucleosynthesis, $g_*=g_s=10.75$. } 
The presence of gravitational waves changes the expansion rate at the big bang nucleosynthesis and at the photon decoupling, affecting the production of light elements and the anisotropy in the cosmic microwave background, similarly to massless neutrinos or dark radiation. 
In addition, fluctuations in the energy density of gravitational waves evolve in the same way as those of neutrinos or dark radiation, and hence their effects on the anisotropy in the cosmic microwave background and structure formation are the same as massless neutrinos or dark radiation. Since in the literature limits on the effective number of neutrino species from the abundance of light elements, cosmic microwave background and structure formation are reported, it is convenient to introduce $N_{\mathrm{eff,GW}}$ by 
writing $\rho_{\mathrm{tot}}$ as 
\begin{equation}
\rho_{\mathrm{tot}}=\frac{\pi^2}{30}\left(2+\frac
{7}{8}\cdot 2\cdot \left(\frac{4}{11}\right)^{4\epsilon/3}[2(1-\epsilon)+N_\nu+N_{\mathrm{eff,GW}}]\right)T^4,
\end{equation}
so that the limits on neutrinos species in the literature can be directly applied to gravitational waves. 
Here, $\epsilon=0$ at the big bang nucleosynthesis and $\epsilon=1$ at the photon decoupling, noting that the electron-positron annihilation takes place between these two epochs, which results in a temperature difference between already-decoupled neutrinos and the rest of radiation. We also set $N_\nu=3+0.046\epsilon$, since neutrinos were not fully decoupled at the electron-positron annihilation \cite{Mangano:2005cc}.
Hence,
\begin{equation}
\frac{\rho_{\mathrm{GW}}}{\rho_{\mathrm{rad}}}\bigg|_{t>t_{\mathrm{th}}}\left
(=\left[\frac{g_*}{g_*(t_{\mathrm{th}})}\right]^{1/3}\left(\frac{\rho_{\mathrm{GW}}}{\rho_{\mathrm{rad}}}\right)\bigg|_{t=t_{\mathrm{th}}}\right)
=g_*^{-1}\cdot\frac{7}{4}\left(\frac{4}{11}\right)^{4\epsilon/3}N_{\mathrm{eff,GW}}.
\end{equation}
Here,
\begin{equation}
g_*=2+\frac{7}{4}\left(\frac{4}{11}\right)^{4\epsilon/3}[2(1-\epsilon)+N_\nu]
\end{equation}
and $g_*=3.38$ and 10.75 at the photon decoupling and the big bang nucleosynthesis, respectively. 
Then, noting Eq. (\ref{relation}), we obtain
\begin{equation}
N_{\mathrm{eff,GW}}=\frac{4}{7}\left(\frac{4}{11}\right)^{-4\epsilon/3}g_*\left(\frac{g_*}{g_*(t_{\mathrm{th}})}\right)^{1/3}\left(\frac{\rho_{\mathrm{GW}}}{\rho_{\mathrm{Higgs}}}\right)\bigg|_{t=t_{\mathrm{late}}}.\label{neffgw}
\end{equation}
The prefactor here turns out to be 2.36 and 2.86 at the photon decoupling and the big bang nucleosynthesis, respectively. 

Examples of $N_{\mathrm{eff,GW}}$ at the photon decoupling for several combinations of the model parameters $(x_0,\xi,\lambda)$ are shown in TABLE 1. 
Note that the result is independent of the energy scale of inflation because both
$\rho_{\mathrm{GW}}$ and $\rho_{\mathrm{Higgs}}$ scales as $H^4$.
The parameters of the numerical calculations include $(d\eta,dk,\eta_i,k_m)$, where $d\eta$ ($dk$) is the interval for $\eta$ ($k$), $\eta_i$ is the initial moment of numerical integration and $k_m$ is the minimum wavenumber. The TABLE I was obtained for $(d\eta,dk,\eta_i,k_m)$=$(0.01,0.01,-10,0.05)$ in unit of the Hubble parameter during inflation, namely, taking $H=1$.  We have obtained almost the same results for $(d\eta,dk,\eta_i,k_m)=(0.02,0.01,-10,0.1),(0.01,0.02,-10,0.1),(0.01,0.01,-20,0.1)$ and $(0.01,0.01,-10,0.1)$. 
Making $x_0$ smaller enhances both spinodal instabilities and gravitational particle production.  Consequently $N_{\mathrm{eff,GW}}$ is not significantly altered by changing $x_0$. Increasing $\xi$ enhances only spinodal instabilities, hence $N_{\mathrm{eff,GW}}$ is smaller for larger $\xi$. TABLE I also shows that the self-interaction term is unimportant for the values of $\xi$ yielding observable $N_{\mathrm{eff,GW}}$, as discussed in the previous section.   
\begin{center}
\begin{table}[t]
    \begin{tabular}{|c|c|c|c|c|c|}
    \hline
    $(x_0,\xi,\lambda)$ & $N_{\mathrm{eff,GW}}$ & $N^{3/4}$ & $(x_0,\xi,\lambda)$ & $N_{\mathrm{eff,GW}}$ & $N^{3/4}$ \\ \hline
    (0.1,1,0.01) &  0.72&5&(0.1,1,0.005)&0.72&5 \\ \hline
    (0.1,2,0.01) & 0.089&23&(0.1,2,0.005)&0.088&23 \\ \hline
    (0.5,1,0.01) & 0.65&5&(0.5,1,0.005)&0.64&5 \\ \hline
    (0.5,2,0.01)&0.067&28&(0.5,2,0.005)&0.064&29\\ \hline
    \end{tabular}
    \captionof{table}{Examples of $N_{\mathrm{eff,GW}}$ at the photon decoupling for several combinations of the model parameters $(x_0,\xi,\lambda)$. The reheating temperature is $4\times 10^{6}N^{3/4}(r/0.01)\mathrm{GeV}$ (see the texts for more details). }
    \end{table}  
\end{center}

The energy density of the Higgs after spinodal instabilities is related to that of gravitons via $\rho_{\mathrm{Higgs}}=(2.86/N_{\mathrm{eff,GW}})\rho_{\mathrm{GW}}=N\rho_r$ from Eq. (\ref{neffgw}), where $\rho_r$ is the energy density of a minimally-coupled massless scalar field created gravitationally provided in Sec. II, $N=5.72/N_{\mathrm{eff,GW}}$ and $N_{\mathrm{eff,GW}}$ is that at the photon decoupling. 
 This means that the reheating temperature is higher by $N^{3/4}$ from Ref. \cite{Artymowski:2017pua}, relative to the reheating temperature given in Sec. II. This factor $N^{3/4}$ is also shown in Table I. 
 
The quantity $N_{\mathrm{eff,GW}}$ can be constrained by the observations of the cosmic microwave background and the structures of the universe \cite{Smith:2006nka,Sendra:2012wh}. 
Future observations of the cosmic microwave background, potentially reaching $\sigma(N_{\mathrm{eff}})\sim 0.02-0.03$ \cite{Abazajian:2016yjj}, or 21 cm line radiation \cite{Oyama:2015gma} would probe larger values of $\xi$ for this reheating scenario. For comparison, there is a 95\% C. L. upper limit on $N_{\mathrm{eff}}^{\mathrm{(upper)}}$ of $4.65$ from Big Bang Nucleosynthesis \cite{Steigman:2012ve,Kuroyanagi:2014nba}. There is also a 95\% C. L. upper limit of 3.77 from Planck power spectra alone \cite{Ade:2015xua}. The dependence of $N_{\mathrm{eff,GW}}$ on $\xi$ is also shown in Fig. 1. 
\begin{figure}[htbp]
  \begin{center}
    \includegraphics[clip,width=7.0cm]{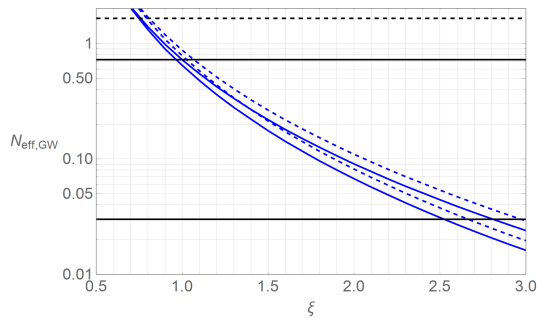}
    \caption{The dependence of $N_{\mathrm{eff,GW}}$ at the photon decoupling (solid) and at the big bang nucleosynthesis (dashed) on the Higgs non-minimal coupling $\xi$. The parameter $x_0$, characterizing the rapidness of the transition from a de-Sitter phase to a kination, is taken as 0.1 (upper) and 0.5 (lower), and $\lambda=0.01$. The horizontal lines from top to bottom show an upper limit from the abundance of the light elements $(4.65-3=1.65)$, the Planck satellite $(3.77-3.046=0.724)$, and a futuristic expected limit of 0.03 (see the text for more detail). }
    \label{fig:hamu}
  \end{center}
\end{figure}

\section{CONCLUSION}

In a class of inflation models in which inflation is followed by a
kination, the universe is supposed to be reheated by gravitational
particle production, which is rather inefficient with a relatively
low reheating temperature. What is worse, gravitons are
created twice as much as a massless minimally coupled bosons, which
 cause problems in big bang nucleosynthesis, CMB observation, 
and cosmic structure formation.  Thus we need additional reheating
mechanism, which may be provided by the standard Higgs field.
However, while condensation of long-wave fluctuations in Higgs field acquired
during inflation can provide an additional source of radiation,
it also causes too large curvature fluctuations eventually, so that
this scenario does not work, either \cite{Kunimitsu:2012xx}.

In this situation, 
we have shown that both of these problems can be solved if the Higgs
field is non-minimally coupled to the scalar curvature, and further
shown that observations which probe extra graviton radiation can be
used to constrain the value of the non-minimal coupling parameter $\xi$
by expressing the energy density of gravitons as an effective, 
additional contribution to the number of neutrino species, denoted by $N_{\mathrm{eff,GW}}$. It is determined by how rapid the transition from inflation to the next phase is ($x_0$), the Higgs non-minimal coupling to gravity $\xi$, and its self coupling $\lambda$, but independent of the energy scale of inflation. 
The values of $N_{\mathrm{eff,GW}}$ for several combinations of these model parameters are presented in TABLE 1. 
It turns out that this quantity is mainly determined by $\xi$, with only
weak dependence on $x_0$. It is hardly affected by changing $\lambda$,
when $\xi$ is not too large and hence $N_{\mathrm{eff,GW}}$ can be
observable.  This indicates that spinodal instability is not so strong
that the Higgs field does not reach the minimum of the potential.

Although the conformal coupling $\xi=1/6$ is sufficient to suppress
the long wavelength fluctuation of Higgs condensation, 
the quantity $N_{\mathrm{eff,GW}}$ is yet too large 
to be consistent with existing observations even for $\xi\simeq 1$, 
so that the graviton creation yields a more stringent lower bound on
$\xi$
than the curvature fluctuations generated from Higgs field.
In other words,  future experiments would be able to 
 probe larger values of $\xi$.

A few comments are in order.
The requirement that the electroweak vacuum be stable \cite{Espinosa:2015qea,Herranen:2015ima,Kohri:2017iyl} after the Higgs growth at the transition would provide additional constraints on this scenario, especially when $\xi$ is large and the substantial spinodal instability is shut off by the self-interaction term, though they would depend on particle physics. 
Gravitational creation of gravitons can also be used to constrain reheating in modified gravity theories. While it is suppressed in $R^2$ inflation \cite{Starobinsky:1980te,Ema:2016hlw}, it constrains an $f(R)$ scenario considered in \cite{Nishizawa:2014zra}. 
Our conclusions would not be significantly altered by inclusion of metric fluctuations, or by a different choice of frame \cite{Markkanen:2017dlc}. 
Our limits are loosened by a late-time entropy production \cite{Nakayama:2008wy,Nakayama:2009ce}. Such a possibility can be explored by a comparison of tensor modes on CMB B-mode polarization scales and on the DECIGO band \cite{Kuroyanagi:2014qza}. 

Though we have restricted our attention to gravitational radiation generated at the transition from a de-Sitter phase to a kination phase, it would be worthwhile to mention gravitational wave frequency spectra $\Omega_{\mathrm{GW}}(f)$ generated {\em during} inflation on different frequencies. The spectrum $\Omega_{\mathrm{GW}}$ behaves as $f^{-2}, f^{0}, f$ for modes which reenter the horizon during  matter domination, a radiation domination and a kination, respectively \cite{Giovannini:1999bh}. The amplitude of the plateau ($\sim f^0$) is roughly given as $\rho_{\mathrm{c}}\Omega_{\mathrm{GW}}\sim f^2h(f)^2\sim f_{\mathrm{eq}}^2H^2(a_0/a_{\mathrm{eq}})^{-2}=f_{\mathrm{R}}^2H^2(a_0/a_{\mathrm{R}})^{-2}$, noting the initial amplitude is $\sim H$ and it decays in proportion to the inverse of the scale factor after the reentry. Here, the subscript R denotes quantities at reheating. The amplitude at $f=f_{\mathrm{K}}$, the comoving frequency corresponding to the beginning of kination, is $\rho_{\mathrm{c}}\Omega_{\mathrm{GW}}(f_{\mathrm{K}})\sim H^4(a_0/a_{\mathrm{K}})^{-4}$, simply because the initial energy density $\sim H^4$ is red-shifted from the moment of generation. If one considers cases where the Universe is reheated through a component created at the same time, the beginning of a kination,  with energy density $H^4$, as in the scenario considered in this paper, then this amplitude just coincides with $\rho_{\mathrm{rad},0}$: $\rho_{\mathrm{c}}\Omega_{\mathrm{GW}}(f_{\mathrm{K}})\sim \rho_{\mathrm{rad},0}$. Furthermore, the frequency is $f_{\mathrm{K}}=H(a_0/a_{\mathrm{K}})^{-1}\sim (c^5\rho_{\mathrm{rad},0}/h)^{1/4}\sim 10^{11}$Hz. Hence, the location and hight of the peak corresponding to the beginning of kination is independent of the Hubble parameter during inflation. (See Fig. 4 of Ref. \cite{Giovannini:1999bh}.)  The energy density of gravitational waves
 corresponding to this peak is somewhat smaller than what we calculated in \S II.
Thus gravitons created from gravitational particle production imposes
more stringent constraint on $\xi$ than tensor perturbations generated
 during inflation.  \\\\\\

 \textbf{Note added }

 After we posted the original version of this paper in the arXiv, Dimopoulos and Markkanen posted a paper
discussing a similar situation \cite{Dimopoulos:2018wfg} there.  They, however, considered the case $\xi$ is much
larger than our case and the scalar field settles to the potential minimum immediately after
the kination commences.

\begin{acknowledgments}
TN thanks Teruaki Suyama and Marc Kamionkowski for helpful input. TN was partially supported by
JSPS Postdoctoral Fellowships for Research Abroad. 
JY was supported by JSPS KAKENHI, Grant-in-Aid for Scientific Research 15H02082 and Grant-in-Aid for Scientific Research on Innovative Areas 15H05888.


\end{acknowledgments}


\begin{thebibliography}{99}

\bibitem{Sato:2015dga} 
 For a review of inflation, see e.g.\  K.~Sato and J.~Yokoyama,
  Int.\ J.\ Mod.\ Phys.\ D {\bf 24}, no. 11, 1530025 (2015).

  
\bibitem{ArmendarizPicon:1999rj} 
  C.~Armendariz-Picon, T.~Damour and V.~F.~Mukhanov,
  Phys.\ Lett.\ B {\bf 458}, 209 (1999)
  [hep-th/9904075].
  
\bibitem{Kobayashi:2010cm} 
  T.~Kobayashi, M.~Yamaguchi and J.~Yokoyama,
  Phys.\ Rev.\ Lett.\  {\bf 105}, 231302 (2010)
  [arXiv:1008.0603 [hep-th]].
  
\bibitem{Kobayashi:2011nu} 
  T.~Kobayashi, M.~Yamaguchi and J.~Yokoyama,
  Prog.\ Theor.\ Phys.\  {\bf 126}, 511 (2011)
  [arXiv:1105.5723 [hep-th]].
  
\bibitem{Peebles:1998qn} 
  P.~J.~E.~Peebles and A.~Vilenkin,
  Phys.\ Rev.\ D {\bf 59}, 063505 (1999)
  [astro-ph/9810509].
  
\bibitem{Ford:1986sy} 
  L.~H.~Ford,
  Phys.\ Rev.\ D {\bf 35}, 2955 (1987).
  
  \bibitem{Starobinsky:1994bd} 
  A.~A.~Starobinsky and J.~Yokoyama,
  Phys.\ Rev.\ D {\bf 50}, 6357 (1994)
  doi:10.1103/PhysRevD.50.6357
  [astro-ph/9407016].
  
\bibitem{Kunimitsu:2012xx} 
  T.~Kunimitsu and J.~Yokoyama,
  Phys.\ Rev.\ D {\bf 86}, 083541 (2012)
  [arXiv:1208.2316 [hep-ph]].
  
\bibitem{Kamada:2014ufa} 
  K.~Kamada,
  Phys.\ Lett.\ B {\bf 742}, 126 (2015)
  [arXiv:1409.5078 [hep-ph]].
 
  
\bibitem{Herranen:2015ima} 
  M.~Herranen, T.~Markkanen, S.~Nurmi and A.~Rajantie,
  Phys.\ Rev.\ Lett.\  {\bf 115}, 241301 (2015)
  [arXiv:1506.04065 [hep-ph]].
 
  
\bibitem{Figueroa:2015rqa} 
  D.~G.~Figueroa, J.~Garcia-Bellido and F.~Torrenti,
  Phys.\ Rev.\ D {\bf 92}, no. 8, 083511 (2015)
  [arXiv:1504.04600 [astro-ph.CO]].
  
\bibitem{Enqvist:2015sua} 
  K.~Enqvist, S.~Nurmi, S.~Rusak and D.~Weir,
  JCAP {\bf 1602}, no. 02, 057 (2016)
  [arXiv:1506.06895 [astro-ph.CO]].
  
\bibitem{Figueroa:2016dsc} 
  D.~G.~Figueroa and C.~T.~Byrnes,
  Phys.\ Lett.\ B {\bf 767}, 272 (2017)
  [arXiv:1604.03905 [hep-ph]].

\bibitem{Cormier:1999ia} 
  D.~Cormier and R.~Holman,
  Phys.\ Rev.\ D {\bf 62}, 023520 (2000)
  [hep-ph/9912483].

\bibitem{Albrecht:2014sea} 
  A.~Albrecht, R.~Holman and B.~J.~Richard,
  Phys.\ Rev.\ Lett.\  {\bf 114}, 171301 (2015)
  [arXiv:1412.6879 [hep-th]].
  
  
  
\bibitem{Smith:2006nka} 
  T.~L.~Smith, E.~Pierpaoli and M.~Kamionkowski,
  Phys.\ Rev.\ Lett.\  {\bf 97}, 021301 (2006)
  [astro-ph/0603144].
\bibitem{Sendra:2012wh} 
  I.~Sendra and T.~L.~Smith,
  Phys.\ Rev.\ D {\bf 85}, 123002 (2012)
  [arXiv:1203.4232 [astro-ph.CO]].
  
\bibitem{Artymowski:2017pua} 
  M.~L.~Artymowski, O.~Czerwinska, Z.~Lalak and M.~Lewicki,
  arXiv:1711.08473 [astro-ph.CO].
  
\bibitem{Lyth:2009zz} 
  D.~H.~Lyth and A.~R.~Liddle,
  Cambridge, UK: Cambridge Univ. Pr. (2009) 497 p
  
 
\bibitem{Bunch:1978yq} 
  T.~S.~Bunch and P.~C.~W.~Davies,
  Proc.\ Roy.\ Soc.\ Lond.\ A {\bf 360}, 117 (1978).
  doi:10.1098/rspa.1978.0060
 
  
\bibitem{Vilenkin:1982wt} 
  A.~Vilenkin and L.~H.~Ford,
  Phys.\ Rev.\ D {\bf 26}, 1231 (1982).
  doi:10.1103/PhysRevD.26.1231
 
  

  
\bibitem{Husdal:2016haj} 
  L.~Husdal,
  Galaxies {\bf 4}, 78 (2016)
  [arXiv:1609.04979 [astro-ph.CO]].
  
\bibitem{Nakama:2016enz} 
  T.~Nakama and T.~Suyama,
  Phys.\ Rev.\ D {\bf 94}, no. 4, 043507 (2016)
  [arXiv:1605.04482 [gr-qc]].
  
\bibitem{Mangano:2005cc} 
  G.~Mangano, G.~Miele, S.~Pastor, T.~Pinto, O.~Pisanti and P.~D.~Serpico,
  Nucl.\ Phys.\ B {\bf 729}, 221 (2005)
  [hep-ph/0506164].
  
   
  
\bibitem{Abazajian:2016yjj} 
  K.~N.~Abazajian {\it et al.} [CMB-S4 Collaboration],
  arXiv:1610.02743 [astro-ph.CO].
\bibitem{Oyama:2015gma} 
  Y.~Oyama, K.~Kohri and M.~Hazumi,
  JCAP {\bf 1602}, no. 02, 008 (2016)
  [arXiv:1510.03806 [astro-ph.CO]].
  
  
\bibitem{Steigman:2012ve} 
  G.~Steigman,
  Adv.\ High Energy Phys.\  {\bf 2012}, 268321 (2012)
  [arXiv:1208.0032 [hep-ph]].

\bibitem{Kuroyanagi:2014nba} 
  S.~Kuroyanagi, T.~Takahashi and S.~Yokoyama,
  JCAP {\bf 1502}, 003 (2015)
  [arXiv:1407.4785 [astro-ph.CO]].
  
\bibitem{Ade:2015xua} 
  P.~A.~R.~Ade {\it et al.} [Planck Collaboration],
  Astron.\ Astrophys.\  {\bf 594}, A13 (2016)
  [arXiv:1502.01589 [astro-ph.CO]].
  
\bibitem{Kohri:2017iyl} 
  K.~Kohri and H.~Matsui,
  arXiv:1704.06884 [hep-ph].






  
  
  
  
  
\bibitem{Espinosa:2015qea} 
  J.~R.~Espinosa, G.~F.~Giudice, E.~Morgante, A.~Riotto, L.~Senatore, A.~Strumia and N.~Tetradis,
  JHEP {\bf 1509}, 174 (2015)
  [arXiv:1505.04825 [hep-ph]].
\bibitem{Herranen:2015ima} 
  M.~Herranen, T.~Markkanen, S.~Nurmi and A.~Rajantie,
  Phys.\ Rev.\ Lett.\  {\bf 115}, 241301 (2015)
  [arXiv:1506.04065 [hep-ph]].
  
  
  
 
  
  
  
  
  
\bibitem{Starobinsky:1980te} 
  A.~A.~Starobinsky,
  Phys.\ Lett.\  {\bf 91B}, 99 (1980).
\bibitem{Ema:2016hlw} 
  Y.~Ema, R.~Jinno, K.~Mukaida and K.~Nakayama,
  Phys.\ Rev.\ D {\bf 94}, no. 6, 063517 (2016)
  [arXiv:1604.08898 [hep-ph]].
  
\bibitem{Nishizawa:2014zra} 
  A.~Nishizawa and H.~Motohashi,
  Phys.\ Rev.\ D {\bf 89}, no. 6, 063541 (2014)
  [arXiv:1401.1023 [astro-ph.CO]].
  
\bibitem{Markkanen:2017dlc} 
  T.~Markkanen, S.~Nurmi and A.~Rajantie,
  arXiv:1707.00866 [hep-ph].
  
\bibitem{Nakayama:2008wy} 
  K.~Nakayama, S.~Saito, Y.~Suwa and J.~Yokoyama,
  JCAP {\bf 0806}, 020 (2008)
  [arXiv:0804.1827 [astro-ph]].
  
\bibitem{Nakayama:2009ce} 
  K.~Nakayama and J.~Yokoyama,
  JCAP {\bf 1001}, 010 (2010)
  [arXiv:0910.0715 [astro-ph.CO]].
  
\bibitem{Kuroyanagi:2014qza} 
  S.~Kuroyanagi, K.~Nakayama and J.~Yokoyama,
  PTEP {\bf 2015}, no. 1, 013E02 (2015)
  [arXiv:1410.6618 [astro-ph.CO]].

\bibitem{Giovannini:1999bh} 
  M.~Giovannini,
  Phys.\ Rev.\ D {\bf 60}, 123511 (1999)
  doi:10.1103/PhysRevD.60.123511
  [astro-ph/9903004].
  
\bibitem{Dimopoulos:2018wfg} 
  K.~Dimopoulos and T.~Markkanen,
  JCAP06(2018)021
  doi:10.1088/1475-7516/2018/06/021
  [arXiv:1803.07399 [gr-qc]].
  
\end{thebibliography}
\end{document}